\documentclass[]{raa}          
\usepackage{graphicx,times}
\usepackage{natbib}
\usepackage{epstopdf}
\usepackage{amssymb,amsmath}

\usepackage[pagebackref=true]{hyperref}
\hypersetup{pdftitle = The title of my PDF, pdfauthor = My name, pdfsubject= The subject, pdfkeywords = keyword1 keyword2 keyword3} 
\hypersetup{colorlinks = true, linkcolor = green, anchorcolor = red, citecolor = blue, filecolor = red, urlcolor = red}

\providecommand{\bm}[1]{\mbox{\boldmath$#1$\unboldmath}}
\def\kms{$\rm{km~s}^{-1}$}

\newcommand{\bahao}{\fontsize{5pt}{\baselineskip}\selectfont}
\def\kms{$\rm{km~s}^{-1}$}  

\def\Ha{{H$\alpha$}}
\def\Hb{{H$\beta$}}
\def\Oiii{[\mbox{O\,{\sc iii}}]}

\def\Sii{[\mbox{S\,{\sc ii}}]}
\def\Nii{[\mbox{N\,{\sc ii}}]}
\def\lo3{\ensuremath{L_{\mathrm{[O\,{\bahao III}]}}}}

\def\apjl{ApJL}                
\def\apjs{ApJS}               
\def\ao{Appl.~Opt.}           
\def\aap{A\&A}                
\def\aaps{A\&AS}              

\begin{document}

   \title{The Broad Wing of \Oiii\ $\lambda$5007 Emission Line in Active Galactic Nuclei}

 \volnopage{ {\bf 20xx} Vol.\ {\bf x} No. {\bf XX}, 000--000}
   \setcounter{page}{1}
 
   \author{Zhixin Peng      \inst{1,2}
   \and Yanmei Chen       \inst{1,2}
   \and Qiusheng Gu       \inst{1,2}
   \and Kai Zhang     \inst{3}
   }
 
   \institute{ School of Astronomy and Space Science, Nanjing University, Nanjing 210093, P. R. China; {\it zxpeng@nju.edu.cn}
\and
Key Laboratory of Modern Astronomy and Astrophysics (Nanjing University), Ministry of Education, Nanjing 210093, China \\
\and
Key Laboratory for Research in Galaxies and Cosmology, Shanghai Astronomical Observatory, Chinese Academy of Sciences, 80 Nandan Road, Shanghai 200030, P. R. China\\
\vs \no
   {\small  Received [year] [month] [day]; accepted [year] [month] [day] }
}
\abstract{ We use a type 2 AGN sample from SDSS DR7 in which the \Oiii\ $\lambda$5007 
emission line can be modeled by two Gaussian components, a broad wing plus a narrow core, to 
investigate the origin of the broad wing and the connection between the velocity 
shift of the broad wing and the physical parameters of active galactic nuclei (AGNs) 
as well as their host galaxies. We find that the flux of the wing components is 
roughly equal to that of the core components in statistic. However, the velocity shift 
of the wing component has only weak, if any, correlations with the physical properties 
of AGNs and the host galaxies such as bolometric luminosity, the Eddington ratio, the mass of 
supermassive black holes, D4000, $\rm H\delta_A$ or stellar mass. Comparing the velocity 
shift from our type 2 AGN sample to that from type 1 sample in \citet{zhang2011} , we suggest 
the \Oiii\ broad wing originates from outflow.
}

   \authorrunning{Z.X. Peng et al. }             
   \titlerunning{Broad Wing  of \Oiii}  
   \maketitle

%
%

\section{Introduction}

Active galactic nuclei (AGNs) feedback now appears as a crucial process in 
galaxy formation and evolution. It is well-known that the tight correlation 
between nuclear black hole mass ($M_{\rm BH}$) and bulge stellar velocity 
dispersion (i.e., the $M_{\rm BH} - \sigma_*$ relation; 
\citealt{ferrarese2000}; \citealt{gebhardt2000}; \citealt{tremaine2002}) is 
compelling evidence for a close connection between the evolution of 
supermassive black holes and their host galaxies. AGN feedback is the most likely 
explanation for this relation (e.g., \citealt{silk1998}; \citealt{fabian1999}). 
\citet{hopkins2006} also indicate the importance of feedback in the 
evolutionary model for starbursts, AGN activity and spheroidal galaxies. 
This feedback can terminate star formation in the host galaxy and cease 
gas accretion onto the nuclear black hole, with the form of radiation, winds, 
jets and outflows. 

The various emission lines from narrow-line regions (NLRs) of AGN are ideally 
suited to study the central regions of AGNs, as well as the interaction between 
the central engine and its host galaxy. Unlike the broad-line 
regions (BLRs), the NLR is spatially resolvable, at least for nearby galaxies. It is 
generally believed that the narrow emission lines are produced by clouds 
illuminated by the central AGN, and the kinematics of the NLR clouds are mainly 
dominated by the gravitational potential of the bulge (e.g.,  \citealt{whittle1992}; 
\citealt{nelson1996}). Since the NLR connects to various factors such as the energy 
input from the central engine, the structure of AGN, radio jets, and star 
formation etc, it accesses a number of key questions of AGN phenomenon.

\Oiii\ $\lambda$$\lambda$4959,5007 emission lines are commonly used to study 
the properties of NLRs. It is usually the strongest narrow line in AGNs at optical 
band and cleanly isolated from other emission and absorption 
features in the optical spectrum. The line profile of \Oiii\ doublets 
in low-redshift AGNs is usually asymmetric. In most case, there is a sharper fall-off 
to the red than to the blue, and the redshift of \Oiii\ is negative compared to 
the system velocity derived from different indicators, such as the low-ionization lines 
(\Nii, \Sii), or stellar absorption lines (\citealt{heckman1981} and consequent researches). 
This asymmetry feature of \Oiii\ line has been suggested as an indicator of 
outflows in Seyfert 1s and 2s (\citealt{heckman1981}; \citealt{whittle1988}; 
\citealt{colbert1996}; \citealt{crenshaw2010}), Narrow 
Line Seyfert 1 galaxies (\citealt{bian2005}; \citealt{komossa2008}), type 1 quasars 
(\citealt{heckman1984}; \citealt{boroson2005}) and narrow line radio galaxies (\citealt{holt2008}).

It is now believed that a broad blue wing in addition 
to the main narrow component is ubiquitous in \Oiii\ 
emission. Many previous studies suggest this blue wing 
is contributed by AGN outflows, they also attempt to correlate 
the parameters of lines, such as the \Oiii\ blueshift 
or/and the \Oiii\ line width, with the physical properties of
AGNs (\citealt{zamanov2002}; \citealt{aoki2005}; \citealt{boroson2005}; 
\citealt{bian2005}; \citealt{komossa2008}). Based on homogeneous samples 
of radio-quiet Seyfert 1 galaxies and QSOs selected from SDSS, 
\citet{zhang2011} find that the blueshift of \Oiii\ has only 
weak correlations with fundamental AGN parameters, such as the nuclear 
continuum luminosity at 5100\AA\ ($L_{5100}$), black hole mass ($M_{\rm BH}$), and 
the Eddington ratio ($L_{\rm bol}/L_{\rm Edd}$). \citet{alexander2012} mentioned 
that in statistics, the width and luminosity of the blue wing increase 
with \Oiii\ luminosity, but independent of radio loudness, indicating 
the outflows are driven by AGN radiation rather than relativistic jets.
\citet{zhang2008} found that Seyfert 1s have lower \Nii/\Ha\ ratios 
than Seyfert 2s and the location of Seyfert 1s on the BPT diagram varies 
with extinction of broad lines, suggesting that the inner dense NLR is 
obscured by dusty torus. The inner dense NLR might be the place where 
the \Oiii\ blue wing originates (\citealt{zhang2013}). \citet{stern2013} revisit the location of 
type 1 and type 2 AGNs on the BPT diagrams, finding similar result as 
\citet{zhang2008}---type 1 AGNs are offset to lower \Sii/\Ha\ and 
\Nii/\Ha\ ratios. However, they conclude that this offset between type 1 
and type 2 AGNs is a selection effect rather than dust extinction.

In this paper, we will explore the asymmetric behavior of 
\Oiii~$\lambda$$\lambda$4959,5007 lines in more detail, studying the origin of \Oiii\ 
asymmetry. In Section 2, we describe the sample selection and data analysis. We 
show the results in Section 3. In Section 4, the origin of the 
broad wing is discussed. Our conclusions are given in Section 5.
Throughout this paper, a cosmology with 
{\slshape H}$_0$ = 70 \kms\ Mpc$^{-1}$, $\Omega_{\rm m} = 0.3$, and 
$\Omega_\Lambda = 0.7$ is adopted.

\section{Sample and Data Analysis}
\label{sect:sample}

\subsection{The Sample}
We begin with the galaxy sample of SDSS (\citealt{york2000}) seventh data release 
(DR7; \citealt{abazajian2009}) and select type 2 active galaxies based on the 
widely used BPT diagram (\citealt{baldwin1981}). The SDSS DR7 spectroscopic 
galaxy catalog contains $\sim$ 930,000 spectra taken through $3\arcsec$ diameter fiber in 
the primary redshift range $0 \la {\rm z} \la 0.3$. Flux and equivalent width 
(EQW) of narrow emission lines (e.g., \Ha, \Nii\ $\lambda$6583, 
\Oiii\ $\lambda$5007, \Hb) as well as line indices D4000,  $\rm H\delta_A$ and 
stellar mass have been publicly available since 2008 in the 
MPA/JHU catalog \footnote{The raw data files of this catalogue can be 
downloaded from http://www.mpa-garching.mpg.de/SDSS/DR7/}. 

The criteria used to select the parent type 2 AGN sample used in our analysis are the following:
\begin{enumerate}
\item { Redshift between $0.01 \leq z \leq 0.3$ and specPrimary $=$ 1. 
The lower redshift limit of 0.01 is applied to avoid the influence of peculiar velocity. 
SpecPrimary $=$ 1 deletes repeat observations from the sample.}
\item { $\log(\Oiii/{\rm H}\beta) > 0.61/[\log(\Nii/{\rm H}\alpha) - 0.47] + 1.19$ 
(the solid curve in Figure 2 from \citealt{kauffmann2003}), or 
$\log(\Nii/{\rm H}\alpha) \geq 0.47$. For those objects with \Ha, \Nii, \Oiii, and \Hb\ 
emission lines detected with signal-to-noise ratio (S/N) $>$ 3, we separate 
type 2 AGNs from other sources using the emission line ratio diagnostics.  }
\item { The EQW of \Oiii\ $\lambda$5007 emission line is smaller than $-$5 
(negative EQW means emission) and the median S/N per pixel in the rest-frame wavelength 
range 4880-4920\AA\ and 5030-5070\AA\ (the continuum around \Oiii) greater than 15. 
The high spectral quality requirement around \Oiii\ region ensures reliable analysis of the line profile.  }
\end{enumerate}
We refer to this sample hereafter as ``parent sample''. It contains 9,389 type 2 
AGNs. Other parameters which would be used in this paper like stellar mass ($M_*$), 
absorption line indices (D4000 and $\rm H\delta_A$) 
and stellar velocity dispersion ($V_{\rm disp}$) are also provided in the MPA/JHU catalog.

\subsection{Fitting the Stellar Continuum}
The aim of this study is to use the \Oiii\ emission line to probe outflows in the NLR. 
In order to get pure emission line spectra, we need to model the stellar continuum 
of each galaxy. As described in \citet{tremonti2004} and \citet{brinchmann2004}, 
the continua and absorption lines of each galaxy are fitted by a stellar 
population. The basic assumption is that any galaxy star formation history 
can be approximated by a sum of discrete bursts. The library of template 
spectra is composed of single stellar population (SSP) models generated 
using a preliminary version of the population synthesis code of 
\citet{charlot2013}, including models of 10 different 
ages (0.005, 0.025, 0.1, 0.2, 0.6, 0.9, 1.4, 2.5, 5, and 10 Gyr) and four 
metallicities (0.004, 0.008, 0.017, and 0.04). For each metallicity, the 
ten template spectra with different ages are convolved to the measured stellar 
velocity dispersion of the SDSS galaxy, and the best-fitting model 
spectrum is constructed from a non-negative linear combination of the ten 
template spectra, with dust attenuation modeled as an additional free parameter. 
The metallicity which yields the minimum $\chi^2$ is selected as the final best 
fit. The fitting results can be found on the SDSS-MPA Web site.

\subsection{Fitting the Emission Lines}
\begin{figure}
\includegraphics[angle=90,width=\textwidth]{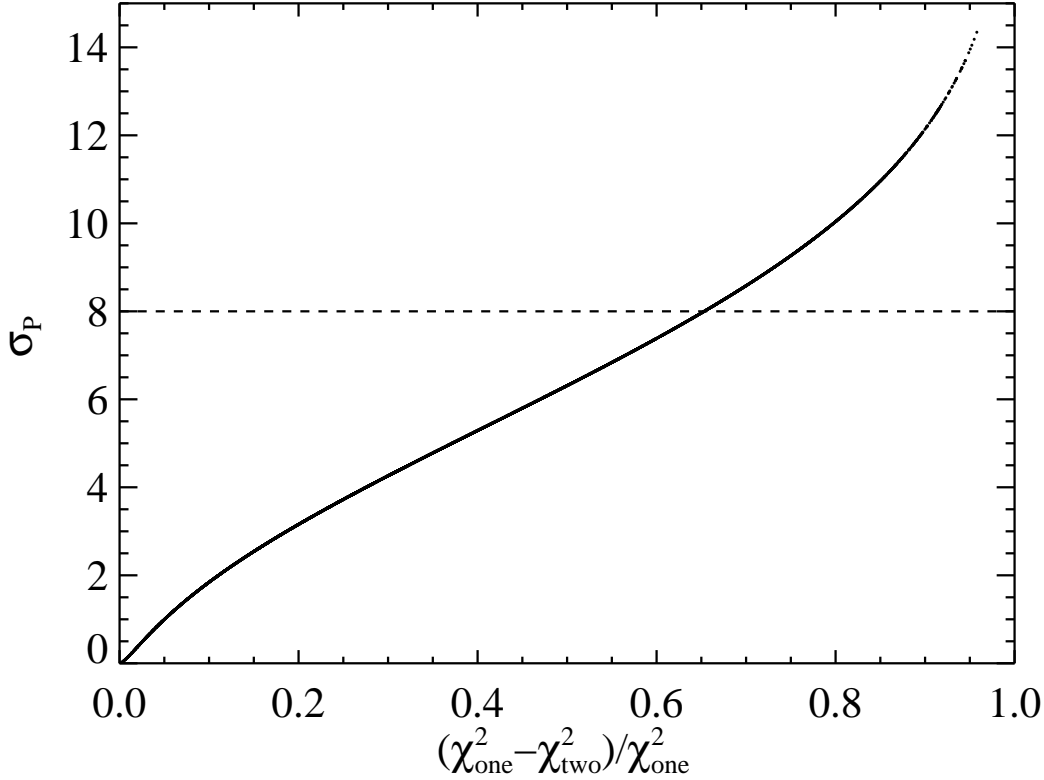}
 \caption{The probability level ($\sigma_{\rm P}$) of the improvement of the fit of emission lines. 
   We select the sources up the horizontal dashed line as our sample.}
  \label{figF-test}
\end{figure}

After subtracting the stellar continuum model, we use the following simple 
method to fit \Oiii\ $\lambda\lambda4959,5007$ emission lines. First, we use 
only one Gaussian to model each \Oiii\ line (hereafter the single-Gaussian model), 
\Oiii\ $\lambda4959$ is forced to have the same profile and shift 
as \Oiii\ $\lambda5007$. We use the galaxy redshift 
from SDSS pipeline to define rest frame. We also decompose each \Oiii\ line 
into two Gaussians (hereafter the double-Gaussian model), a narrow core 
($\Oiii_{4959}^{\rm NC}\ \mbox{and}\ \Oiii_{5007}^{\rm NC}$) 
and a broad wing 
($\Oiii_{4959}^{\rm BW}\ \mbox{and}\ \Oiii_{5007}^{\rm BW}$). 
Each component of 
\Oiii\ $\lambda4959$ is tied to the relevant 
component of \Oiii\ $\lambda5007$ in the same way as that in the single-Gaussian 
model. The line center is limited in the range of 4980-5050 \AA. We compare the reduced $\chi^2$ of the single-Gaussian and 
double-Gaussian model, and use the F-test \citep[chap. 12.1]{lupton1993} to 
calculate how significantly the fit is improved by the double-Gaussian model. 
Figure \ref{figF-test} shows the probability level ($\sigma_{\rm P}$) that 
the double-Gaussian model can improve the fit of emission lines, as a function of the improvement 
of $\chi^2$, which is defined as 
$(\chi^2_{\rm one}-\chi^2_{\rm two})/\chi^2_{\rm one}$. 
$\chi^2_{\rm one}$ is the reduced $\chi^2$ of the single-Gaussian model and $\chi^2_{\rm two}$ 
is the reduced $\chi^2$ of the double Gaussian model. We select the 1,630 sources up 
the horizontal dashed line as our sample for studying outflow. These galaxies require two 
Gaussians at a significance greater than 8$\sigma$, with $\chi^2$ improvement 
greater than $\sim$65\%. 

\section{Results}
\label{sect:results}
\begin{figure}
\includegraphics[angle=0,width=\textwidth]{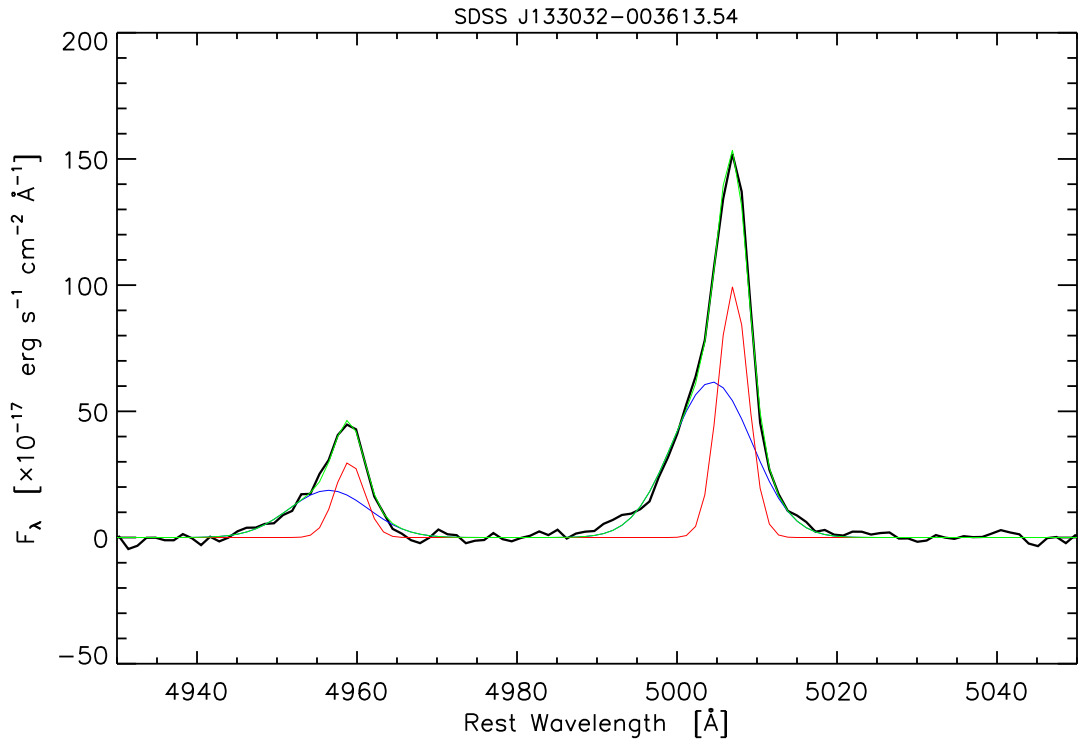}
   \caption{ An example of double Gaussian fit. For each \Oiii\ emission line, two Gaussian 
   components are used. The red represents the core component, and the blue one is the 
   underlying broad wing. The best fit model is shown in green. }
   \label{figexample}
\end{figure}

With the double-Gaussian model fitting of the 1,630 sources in the sample, 
we find the velocity shifts of the core component, 
$V_{\rm off}^{\rm core} = (\lambda_{\rm core} - \lambda_0)/\lambda_0 \times c$, 
has a Gaussian 
distribution over a range of $-200 \sim 200$ \kms, with a median value at 8 \kms\ 
and 68\% of the total probability distribution distributed over 
the range $-37 \sim 53$ \kms. Here $\lambda_{\rm core}$ is the central wavelength 
of the core component, $\lambda_0 = 5006.84$ is the rest-frame line center of \Oiii\ in 
air, $c$ is the speed of light. The distribution of 
velocity shifts of the wing component is highly deviated from Gaussian with a 
median shift of $-$72 \kms. We note that the pipeline redshift is determined 
from both the emission lines and the continuum. If the emission lines are 
blueshifted, then the pipeline redshifts tend to be underestimated. Here we re-determine 
the system redshift from the continuum and absorption lines: starting from the 
stellar continuum model given in section 2.2, we iteratively increase/decrease 
the system velocity by 5 \kms\ and re-calculate a $\chi^2$ value, the absorption 
line redshift is determined by the case with the lowest $\chi^2$ value. In the 
following section, we use this absorption line redshift to define rest-frame. 
Figure \ref{figexample} shows 
one example of the two component fit, the black is the observed emission line spectrum, 
the broad wing and the narrow core are shown in blue and red, respectively, 
the best fit model is over-plotted in green.

\subsection{Correlation between Wing and Core Flux}
\begin{figure}
\includegraphics[angle=90,width=\textwidth]{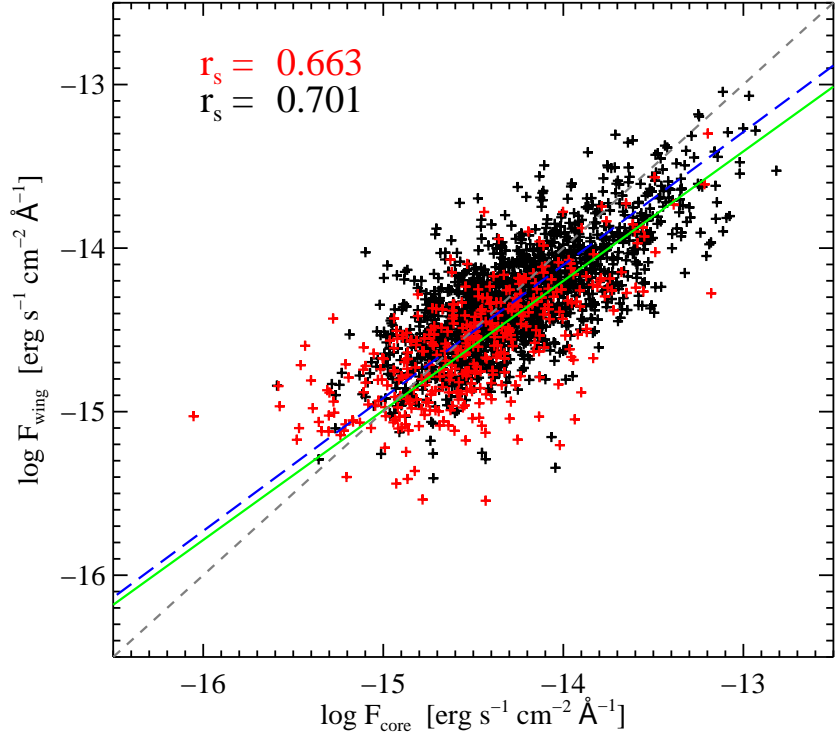}
   \caption{ Correlation between the fluxes of broad wing and narrow core 
   of \Oiii\ $\lambda$5007 emission line in type 1 ($red$) and type 2 ($black$) AGNs. 
   The Spearman rank-order correlation coefficients ($r_{\rm s}$ ) are 0.663 (0.701) 
   for type 1 (type 2) AGNs. The best linear Least-squares approximation is shown, 
   green line for type 1s, and blue long dash line for type 2s. 
    The gray dash line is the x = y line. }
   \label{figc_w}
\end{figure}

In Figure \ref{figc_w}, we show the correlation between the fluxes of the wing ($F_{\rm wing}$) and 
core ($F_{\rm core}$) of \Oiii\ $\lambda5007$ emission line for both type 1 (red dots) and 
type 2 (black dots) AGNs sample. The type 1 AGN sample contains 383 objects from 
\citet{zhang2011} with a redshift range of $0.01\leq z \leq 0.3$. The green line, 
${\rm \log}F_{\rm wing} = (0.792\pm0.070) ~ {\rm \log}F_{\rm core} - (3.112\pm1.016)$, 
is the best linear least-square fit for type 1 AGNs with the Spearman rank-order 
correlation coefficient ($r_{\rm S}$) of 0.663, while the blue line, 
${\rm \log}F_{\rm wing} = (0.724\pm0.035) ~ {\rm \log}F_{\rm core} - (3.964\pm0.496)$, 
is the fit for type 2 AGNs, with $r_{\rm S} = 0.701$. 
\citet{zhang2011} found that on average, the core component comprise 54\% 
of the total emission, which is consistent with a 52\% contribution of 
the core component in our type 2 sample. The detailed explanation of this strong 
correlation between the fluxes of core and wing components in both type 1 and type 2 AGNs 
is beyond the scope of the current paper, we will build a model to 
understand this correlation in a following paper.

\subsection{How $V_{\rm off}^{\rm wing}$ Depends on the Properties of Galaxies}
Since the blue asymmetry of \Oiii\ profile generally suggests the exist of outflow, we explore 
whether the shift of the broad wing is connected with the strength of AGN activity and star 
formation which are the primary driver of the outflow.
We estimate the bolometric luminosity of the AGN as 
$L_{\rm bol} \approx 600\ \lo3$ (\citealt{kauffmann2009}). 
\lo3\ is the total luminosity of wing and core with dust extinction from Balmer decrement, 
and the correlation in this section is independent on which \Oiii\ luminosity we use, 
$\lo3^{\rm wing}$, $\lo3^{\rm core}$, or $\lo3^{\rm wing} + \lo3^{\rm core}$. The 
Eddington ratio is defined as 
$\lambda = L_{\rm bol}/L_{\rm Edd}$, where 
$L_{\rm Edd} \equiv 1.26~\times~10^{38}~ (M_{\rm BH}/M_\odot)~\rm{erg~s}^{-1}$. 
The mass of the central black hole is 
estimated from the famous $M_{\rm BH} - \sigma_*$ relation 
(\citealt{ferrarese2000}; \citealt{gebhardt2000}) of the form 
$\log({M_{\rm BH}}/{M_\odot}) = 8.13 + 4.02 \log({\sigma_*}/{200})$ 
(\citealt{tremaine2002}), where $\sigma_*$ is the stellar velocity dispersion. 

\begin{figure*}
\begin{center}
\includegraphics[angle=90,width=\textwidth]{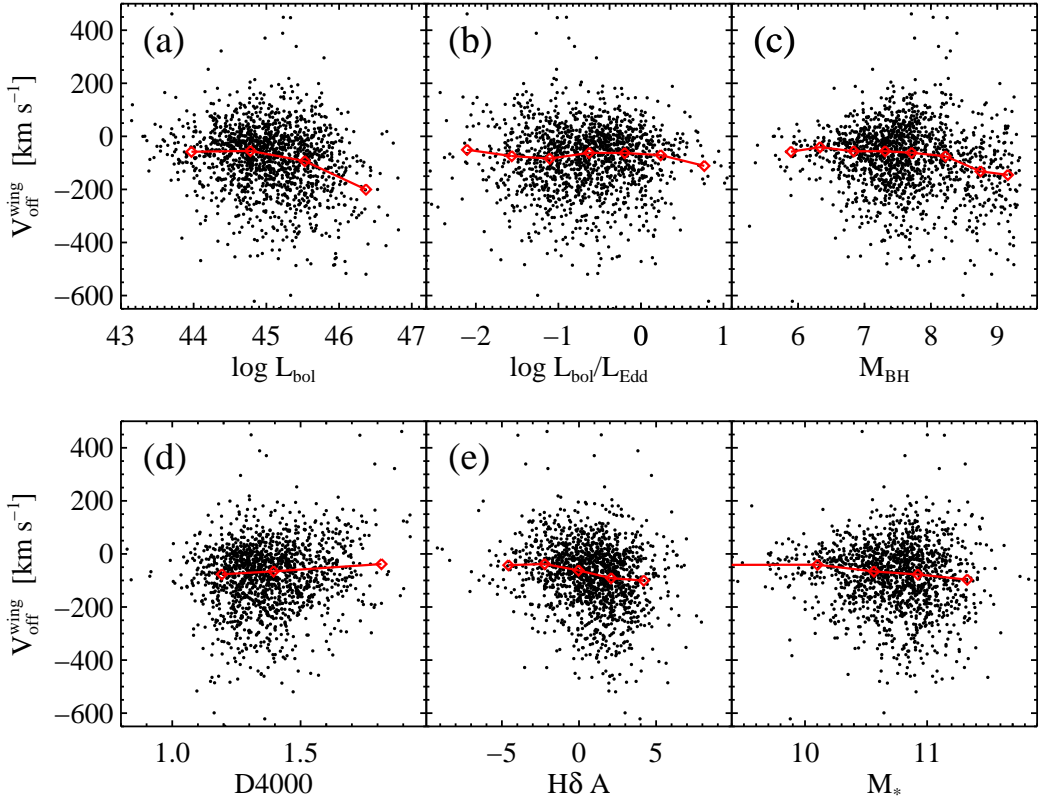}
\end{center}
   \caption{ Velocity shifts of the wing component relative to the system velocity 
   $V_{\rm off}^{\rm wing}$ versus $L_{\rm bol}$, $\lambda$, $M_{\rm BH}$, D4000, 
   $\rm H\delta_A$ and $M_*$. Diamonds are the position of median value in each bin. }
   \label{figproperty}
\end{figure*}

Figure \ref{figproperty} shows the velocity shift of 
wing component ($V_{\rm off}^{\rm wing}$) as a function of $L_{\rm bol}$, 
$\lambda$, mass of supermassive black holes ($M_{\rm BH}$), D4000, 
$\rm H\delta_A$ and stellar mass ($M_*$). 
The stellar mass ($M_*$) and absorption line indices (D4000 and $\rm H\delta_A$) 
are provided in MPA/JHU catalog. The red lines are the median.
The correlation results are listed in Table \ref{tab-property}, where $r_{\rm S}$ is 
the Spearman rank-order correlation coefficient and $P_{\rm null}$ is the probability 
for the null hypothesis of no correlation. The high significant of $P_{\rm null}$ 
is due to the large number of sources in our sample. 
In summary, we have not found any distinct correlation between 
$V_{\rm off}^{\rm wing}$ and galaxy properties. Outflow is driven by both AGN and star 
formation activity. However, the contribution of AGN and star formation activity to the outflow 
varies from object to object. This leads to a lack of correlation in Figure \ref{figproperty}. The result 
is consistent with that in \citet{komossa2008} and \citet{zhang2011}.

\setcounter{table}{0}
\begin{table*}
\begin{minipage}[c]{\textwidth} 
\caption{Correlation Results between $V_{\rm off}^{\rm wing}$ and Galaxy Properties}
\label{tab-property}
\small
\begin{center}
\begin{tabular}{lcccccc}
\\
\hline\hline
correlation & $L_{\rm bol}$ & $\lambda$ & $M_{\rm BH}$ & D4000 & $\rm H\delta_A$ & $M_*$  \\
  \hline\noalign{\smallskip}
$r_{\rm S}$ & -0.184 & 0.010 & -0.083 & 0.097 & -0.204 &  -0.083 \\
$P_{\rm null}$ & 8.9e-13  &  6.8e-1 & 1.3e-3  & 1.9e-4  & 1.9e-15 & 1.3e-3 \\
\hline\hline
\end{tabular}
\end{center}
\end{minipage}
\end{table*}

\subsection{Comparison with Type 1 AGNs}
\begin{figure}
\includegraphics[angle=90,width=\textwidth]{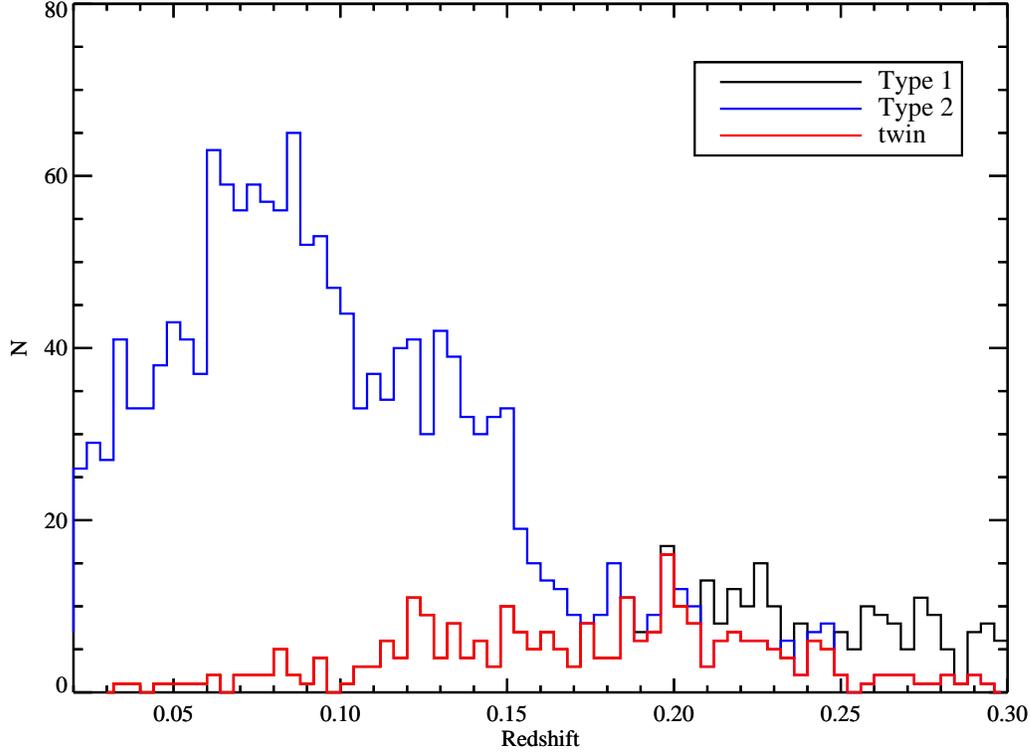}
   \caption{Redshift distribution of  type 1 ($black$) and type 2 ($blue$) AGN sample. 
   The red histogram shows the redshift distribution of the twin sample.}
   \label{figz}
\end{figure}

\begin{figure}
\includegraphics[angle=90,width=\textwidth]{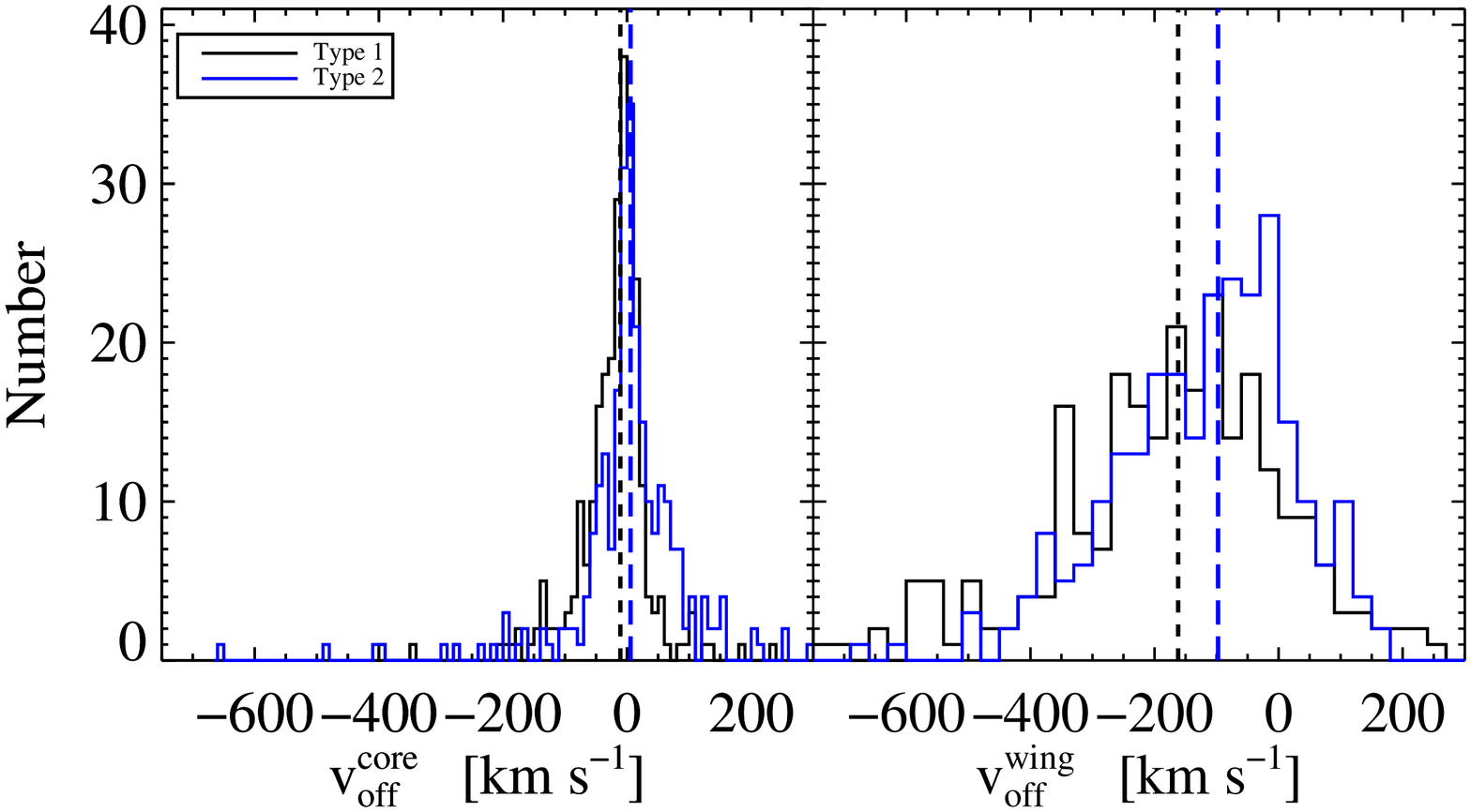}
\caption{ Distributions of velocity offset relative to the system redshift for type 1 ($black$) 
and type 2 ($blue$) twin samples. The left panel for the core component, while the 
right panel for the wing component. The median values are marked by black vertical  
dash line for type 1s and blue vertical long dash line for type 2s. 
 }
\label{figvoff}
\end{figure}

\begin{figure}
\includegraphics[angle=90,width=\textwidth]{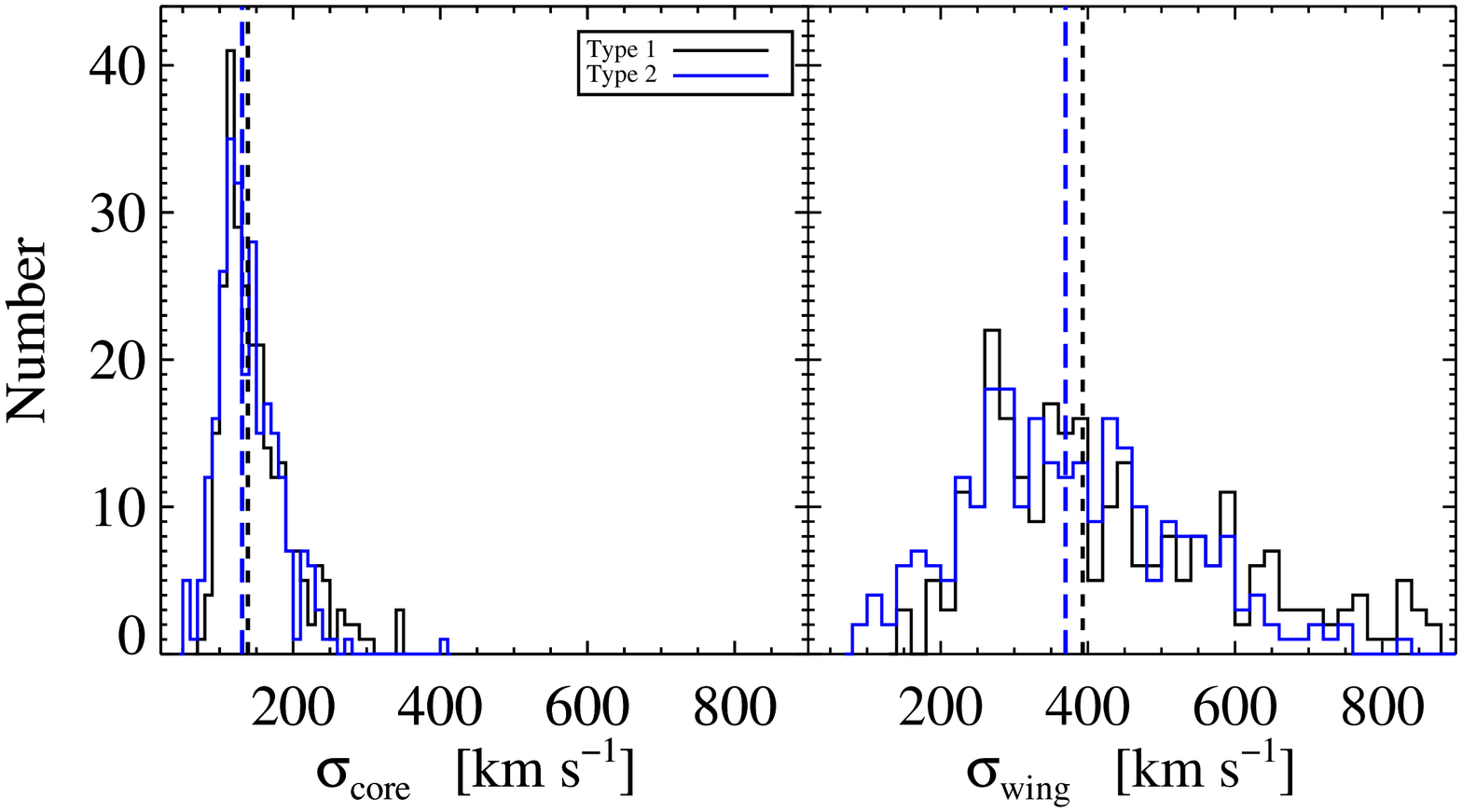}
\caption{ Distributions of the line width $\sigma$ for type 1 ($black$) and type 2 ($blue$) 
twin samples. The left panel for the core component, while the right panel for the wing 
component. The median values are marked by black vertical  
dash line for type 1s and blue vertical long dash line for type 2s.
}
\label{figsig}
\end{figure}

In the standard AGN unified scheme, type 1 and type 2 AGNs are intrinsically 
the same objects. As a result of obscuration of torus, the radiation originated 
from the accretion disc will push materials out mostly along the rotation axis of the 
accretion disc. If the blue wing of \Oiii\ is triggered by outflows, we 
should observe a faster velocity in type 1 than in type 2 AGNs due to the inclination 
effect. In order to avoid 
any evolution effect and make a fair comparison between type 1 and 
type 2 AGNs, we construct twin subsamples from the type 1 (\citealt{zhang2011}) 
and type 2 AGNs by matching their redshift with a tolerance 
of $\Delta z$ = 0.004, namely, the type 1 and type 2 subsamples have exactly the 
same redshift distribution. See the histograms in Figure \ref{figz}. The black, blue and 
red lines show the redshift distributions for type 1, type 2 AGNs, and the matched 
twin sample, respectively. Through this 
redshift match, each subsample contains 264 objects. Figure \ref{figvoff} $\sim$ \ref{figsig} show 
the distributions of velocity offset and line width of the \Oiii\ wing and 
core components for the twin subsamples, and the 
median values are shown by vertical 
dashed lines, black for type 1 and blue for type 2 AGNs.

In Fig. \ref{figvoff}, we show the distributions of the velocity offset relative 
to the system velocity, $V_{\rm off}^{\rm core}$ and $V_{\rm off}^{\rm wing}$. The system 
velocity is derived from \Sii\ emission line for type 1 sample and from stellar 
absorption lines for type 2 sample. The median values of $V_{\rm off}^{\rm core}$ are 
$-11$ \kms, and 6 \kms\ for type 1 and type 2 subsamples, and the median values 
of $V_{\rm off}^{\rm wing}$ are $-162$ \kms, and $-97$ \kms\ for type 1 and type 2 
subsamples, respectively.

In Fig. \ref{figsig}, we show the distributions of line width of wing ($\sigma_{\rm wing}$) and 
core ($\sigma_{\rm core}$) components. The median 
values of $\sigma_{\rm core}$ are 138 \kms\ and 131 \kms\ for 
type 1 and type 2 subsamples, and the median values of $\sigma_{\rm wing}$ are 393 \kms\ and 
370 \kms\ for type 1 and type 2 subsamples, respectively. Basically, there is no 
difference in $\sigma_{\rm core}$ and $\sigma_{\rm wing}$ between type 1 and type 2 
AGNs.

\section{Origin of the Broad Wing}
In this section, we discuss the origin of the broad wing of \Oiii\ $\lambda\lambda4959,5007$, 
including its location and 
physical mechanism, based on the observational results in section 3.
\begin{enumerate}
\item { \textsl{\textbf{Location}}. We derive black hole mass ($M_{\rm BH}$) from the 
$M_{\rm BH}-\sigma_*$ relation. At the same time, if we assume the region which generates the wing 
component is still dominated by the potential of central super-massive black hole, we can 
estimate the location of the region ($R_{\rm wing}$) where the wing comes from as 
$R_{\rm wing} = G \frac{M_{\rm BH}}{f \Delta V^2}$, $f$ is the scaling factor with 
a value of 3.85 (\citealt{collin2006}), $\Delta V$ is the emission line width, we set 
$\Delta V = \sigma_{\rm wing}$, 
$G = 6.67384 \times 10^{-11}\ {\rm m}^3\ {\rm kg}^{-1}\ {\rm s}^{-2}$ is gravitational constant. 
Finally, we get $R_{\rm wing}$ with a median value of 
ten pc for both type 1 and type 2 subsamples. 
We stress that the value of $R_{\rm wing}$ we derive should be a lower limit
since the region that the blue wing originates is not virialized. 
}
\item { \textsl{\textbf{Physical mechanism}}. In section 3.3, we find the wing 
component has a median of $V_{\rm off}^{\rm wing} = -162\ (-97)$ \kms\ for the 
type 1 (type 2) subsample. If we assume the velocity offset of wing component 
originates from the outflows which blow 
out in a direction perpendicular to the accretion disk, we would expect 
$V_{\rm off}^{\rm wing}$(type 2) = $V_{\rm off}^{\rm wing}$(type 1) $\times \cos \theta$, 
where $\theta$ is the opening angle of torus, $V_{\rm off}^{\rm wing}$(type 1) and 
$V_{\rm off}^{\rm wing}$(type 2) are median 
values of the velocity offset of the wing component for type 1 and type 2 AGNs, 
respectively. Applying $V_{\rm off}^{\rm wing}$(type 1) = $-$162 and 
$V_{\rm off}^{\rm wing}$(type 2) = $-$97, we get $\theta \sim 50 \degr$, 
this result is consistent with that from literatures (e.g., \citealt{netzer1987}, \citealt{krolik1994}).}
\end{enumerate}

In addition, we have not found any distinct correlation between 
$V_{\rm off}^{\rm wing}$ and galaxy properties (D4000, $\rm H\delta_A$, stellar mass), as well as the physical 
properties of AGNs (bolometric luminosity, the Eddington ratio, black hole mass). The result 
is consistent with several previous studies (e.g., \citealt{komossa2008}; \citealt{zhang2011}). 
If we accept a scenario in which the low-velocity gas in the core component is dominated by the gravity of the bulge, while the wing is more strongly influenced by the outflow cloud of active nucleus (e.g., \citealt{zamanov2002}; \citealt{greene2005}), the terminal outflow velocity would depend on the origin of the outflow on the one hand, and the deceleration mechanism on the other hand. \citet{komossa2008} discussed several possibilities to explain the acceleration and entrainment of the NLR outflow, including radiation pressure, entrainment in radio jets, thermal winds, and high Eddington ratio. So the launching velocity of the outflow cloud is determined by different acceleration mechanisms and/or different stages of the AGN activity. Meanwhile, NLR outflow is decelerated by the ISM of the host galaxy. Denser ISM results in more efficient deceleration, implying lower velocity (\citealt{zhang2011}). Anyway, both accelerated mechanisms and the column density of NLR cloud would lead to different terminal velocities, thereby explaining the lack of correlation between the observed $V_{\rm off}^{\rm wing}$ and the physical properties of AGNs. 

\section{Conclusion}
We select a type 2 AGN sample from SDSS DR7. In this sample, two Gaussian components 
are required to model the \Oiii\ $\lambda$5007 emission line, a broad wing plus a narrow core.
We measure the velocity shift (relative to the absorption lines), line width and flux of both 
components. Combining our type 2 AGN sample with a type 1 sample from \citet{zhang2011}, 
we find that:
\begin{enumerate}
\item {there is a tight correlation between the fluxes of wing and core components in  both type 1 
and type 2 samples. In both samples, the flux of the wing components is roughly equal to that 
of the core components.}
\item {in the unification scheme of AGNs, the type 1 and 2 AGNs are intrinsically the same, their 
different appearance is due to that we observe them from different direction, a dusty torus blocks 
the continuum source and broad-line region in type 2 AGNs. The difference in the velocity shift 
of the broad wing between type 1 and type 2 AGNs consists with a picture in which the broad wing 
originates from outflows and the outflows blow out in a direction perpendicular to the 
accretion disk with a certain opening angle.}
\item {the velocity shift of the wing component has only weak, if any, correlations with the physical 
properties of AGNs (bolometric luminosity, the Eddington ratio and the mass of supermassive 
black holes) and the host galaxies (D4000, $\rm H\delta_A$ or stellar mass). 
We suggest the lack of correlation is due to that the outflow is driven by both AGN and star 
formation activity. However, the contribution of AGN and star formation activity to the outflow 
varies from object to object. Future IFU survey like MaNGA (Mapping Nearby Galaxies at APO) 
will help us to understand, AGN and star formation, which is the primarily driver of outflow in a certain object.}
\end{enumerate}

\normalem
\begin{acknowledgements}
Zhixin Peng thanks Jing Wang for useful discussions.
The research is supported by the National Natural Science Foundation 
of China (NSFC; Grant Nos. 11273015, 11133001 and 11003007), 
the National Basic Research Program  (973 program No. 2013CB834905), 
and Specialized Research Fund for the Doctoral Program of Higher Education 
(20100091110009).

Funding for the creation and distribution of the SDSS Archive has
been provided by the Alfred P. Sloan Foundation, the Participating
Institutions, the National Aeronautics and Space Administration, the
National Science Foundation, the US Department of Energy, the
Japanese Monbukagakusho, and the Max Planck Society. The SDSS Web
site is \url{http://www.sdss.org}.

The SDSS is managed by the Astrophysical Research Consortium (ARC)
for the Participating Institutions. The Participating Institutions
are The University of Chicago, Fermilab, the Institute for Advanced
Study, the Japan Participation Group, The Johns Hopkins University,
Los Alamos National Laboratory, the Max- Planck-Institute for Astronomy (MPIA), the Max-Planck-Institute for Astrophysics (MPA),
New Mexico State University, University of Pittsburgh, Princeton
University, the United States Naval Observatory and the University
of Washington.
\end{acknowledgements}



\bibliographystyle{raa}
\bibliography{pengbib}

\label{lastpage}

\end{document}